\documentclass{ws-procs9x6}

\begin{document}

\title{Atomic Hydrogen Cleaning of Polarized G\lowercase{a}A\lowercase{s} Photocathodes\footnote{\uppercase{T}his work was supported in part by the \uppercase{U.~S.~D}epartment of \uppercase{E}nergy under contract numbers \uppercase{DE-AC}02-76\uppercase{SF}00515 (\uppercase{SLAC}), and \uppercase{DE-AC}02-76\uppercase{ER}00881 (\uppercase{UW}).}}

\author{T. Maruyama$^\dagger$, D.-A. Luh$^\dagger$, A. Brachmann$^\dagger$,
J.~E. Clendenin$^\dagger$, E.~L. Garwin$^\dagger$,
S. Harvey$^\dagger$, R.~E. Kirby$^\dagger$, 
C.~Y. Prescott$^\dagger$, \\
and R. Prepost$^\ddagger$}
\address{
$^\dagger$Stanford Linear Accelerator Center, Menlo Park, California~94025 \\
$^\ddagger$Department of Physics, University of Wisconsin, Madison, Wisconsin~53706}

\maketitle

\abstracts{
Atomic hydrogen cleaning followed by heat cleaning at 450$^\circ$C
was used to prepare negative-electron-affinity GaAs photocathodes.
When hydrogen ions were eliminated, quantum efficiencies of 15\% were obtained
for bulk GaAs cathodes, higher than the results obtained using conventional
600$^\circ$C heat cleaning.
The low-temperature cleaning technique was successfully applied to thin,
strained GaAs cathodes used for producing highly polarized electrons.
No depolarization was observed even when the optimum cleaning time of
about 30 seconds was extended by a factor of 100.
}

\section{Introduction}
Recently the high-gradient-doping technique has been applied to
photocathode structures to successfully overcome the surface charge limit
effect while maintaining high spin polarization.
The high-gradient-doping technique consists of a thin (10 nm),
very-highly-doped (5$\times$10$^{19}$ cm$^{-3}$) surface layer with
a lower density doping (5$\times$10$^{17}$ cm$^{-3}$) in the remaining
active layer.
High dopant density promotes recombination of the minority carriers trapped
at the surface. Trapped carriers increase the surface barrier in proportion
to the arrival
rate of photoexcited conduction band (CB) electrons. Because CB electrons
depolarize as they diffuse to the surface of heavily doped materials,
the highly doped layer must be very thin, typically no more than a few
nanometers.
However, to achieve high quantum efficiencies, an negative-alectron-affinity (NEA) surface is required,
which in turn must be prepared on an atomically clean surface.
The conventional way to achieve a surface free of all surface oxides and
carbon-related contaminants is to heat the crystal to 600$^\circ$C for
about 1 hour. After only about 2 hours at this temperature, diffusion of
the dopant in the thin,
highly-doped layer results in the re-appearance of the surface charge limit
effect.
Therefore, high temperature heat cleaning should be avoided.

Atomic hydrogen cleaning (AHC) is a well known technique for removing oxides
and carbon-related contaminants at relatively low temperatures.
While As-oxides and Ga$_2$O-like oxides are liberated at temperatures less
than 450$^\circ$C, the removal of Ga$_2$O$_3$-like oxides requires a higher
temperature. Under atomic hydrogen irradiation, Ga$_2$O$_3$-like oxides are
converted to more volatile Ga$_2$O-like oxides.
On the other hand, it has been well demonstrated that atomic hydrogen can
passivate both shallow donor and acceptor impurities.
The passivation rate increases rapidly with the doping concentration.
Since the band-bending in the photocathode is controlled by the $p-$type
doping, the dopant passivation may have an adverse effect on QE.

In the present paper, a systematic study of AHC
in a vacuum-loading system is reported. The AHC system and the
associated analysis system
remain under UHV, while the sample is introduced in the UHV system
through a loading chamber, and transferred between the AHC and analysis
systems under vacuum \cite{tm}.

\section{Experiment}
Two types of GaAs samples were used. Samples (13 $\times$ 13 mm$^2$) cut
from Zn-doped (1$\times$10$^{19}$ cm$^{-3}$) bulk GaAs (001) wafers were used
for optimizing the AHC conditions. Strained GaAs samples
with the active 100-nm GaAs layer Zn-doped at 5$\times$10$^{18}$ cm$^{-3}$
were used for studying the AHC effect on polarization.
Prior to installation
in the loading chamber, a sample was degreased in a boiling solution of
trichloroethylene and
chemically cleaned in NH$_4$OH to remove native oxides on the surface.
Since NH$_4$OH etches only oxides without disturbing the stoichiometry of GaAs,
it was used for the epitaxial photocathodes as well.
Some samples were installed without the NH$_4$OH cleaning
to intentionally leave native oxides on the surface.

The experiments were carried out in a three-chamber UHV
system consisting of a loading chamber, an AHC system and an analysis system
called Cathode Test System (CTS).
The AHC system was equipped with an rf plasma discharge source, a heater tower,
and a linear translator. The surface temperature
during AHC was maintained at 300 - 350$^\circ$C.
The heater tower was electrically isolated from the AHC system so that
a bias voltage could be applied to the GaAs sample during AHC.
Atomic hydrogen was produced by dissociating molecular hydrogen in
a 2.5 cm diameter Pyrex glassware surrounded by a helical rf resonator
following the design used at Jefferson Lab. \cite{sinclair}
To study the effect of hydrogen ions generated by the rf dissociator,
the GaAs sample could be
biased negatively to enhance ion collection. To reduce
the ion current, a transverse magnetic field was applied at the exit of the
dissociator using a permanent magnet.
With the magnet in place, the ion current was negligible ( $<$ 1 nA).

Activation to an NEA surface, and measurement of
QE and polarization were made in the CTS.
After AHC, the sample was transferred to the CTS under vacuum
when the AHC system pressure dropped
to a few 10$^{-8}$ Torr, typically within 30 minutes after AHC.
The cathode activation method used to obtain an NEA surface
consisted of heat cleaning
for 1 hour, cool-down for an hour, followed by
application of cesium until the photo-yield peaked, and then cesium and
nitrogen-trifluoride co-deposition until the photo-yield was again maximized.

AHC was performed on NH$_4$OH-cleaned bulk GaAs samples under three
different conditions that control the ion current: 1)
with no bias and no magnet, 2) with a negative 88-Volt bias and no magnet to enhance
the ion effect, and 3) with magnet and no bias to eliminate the ion effect.
For each AHC, a fresh cathode sample was used and only the AHC time was
varied while all
other conditions were fixed. After AHC, the sample was transferred to the CTS
under vacuum,
heat-cleaned at 450$^\circ$C, activated, and the QE measured.
Figure 1 shows the QE at 670 nm as a function of the AHC time.
QEs as high as 16\% were obtained with only 15 - 30 seconds of AHC.
The QE decreased with prolonged cleaning, yielding only 1.8\% after 40 minutes
of AHC. The QE was higher when the ion current was eliminated
using the magnet, and lower when the ion current was enhanced using the
negative bias.
The data indicate that the excessive absorption of atomic hydrogen in GaAs
is detrimental for the QE.
Under atomic hydrogen irradiation, the native oxides on the surface are
converted to more volatile oxides and get liberated.
If the irradiation continues, atomic hydrogen is absorbed in the GaAs,
passivating the $p$-type dopants in the band bending region.
Since the doping concentration at the surface controls
the band bending, dopant passivation raises the vacuum level,
resulting in a lower QE.
Acceptor passivation by hydrogen proceeds through ion pair formation of
negatively charged acceptors and positively charged hydrogen ions.
\cite{pearton}
Ions are more effective than atomic hydrogen at passivating the dopants.

AHC was also performed on bulk GaAs samples without the NH$_4$OH cleaning.
QEs as high as 14\% were obtained after 1 hour of AHC
(open square in Figure 1).
The sample with native oxides produced a high QE after prolonged AHC,
indicating the oxide layer was protecting the GaAs surface from impinging
atomic hydrogen. \cite{ide}
This indicates the AHC time must be optimized depending on the
oxide level on the GaAs surface.
The NH$_4$OH etching establishes a reproducible level of native oxides.

\begin{figure}[ht]
\centerline{\epsfxsize=2.5in\epsfbox{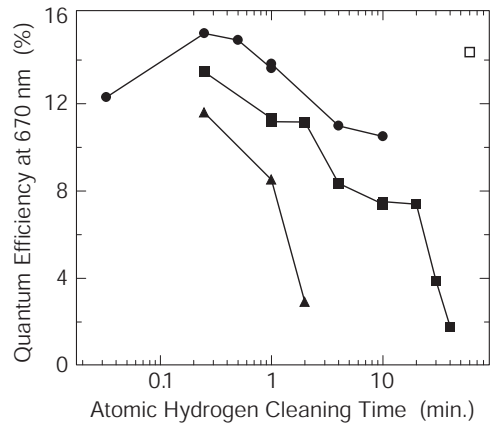}}
\caption{QE at 670 nm as a function of the AHC time. Three different
conditions are used to control ions: solid circles are no bias and with
magnet, squares are no bias and without magnet, and triangles are
with -88 V negative bias without magnet. One sample (open square) was
not cleaned in NH$_4$OH.}
\end{figure}

Studies at Jefferson Lab \cite{baylac} indicate that a significant
depolarization may occur as a result of long exposures to atomic deuterium.
The electron polarization was measured in the present experiment
as a function of the AHC time using thin strained photocathodes.
The sample was heat-cleaned at 570$^\circ$C without AHC and the polarization
was measured. Then, a sequence consisting of 15 minutes of AHC followed by
polarization measurements was repeated four times for the same sample.
The AHC was performed with the magnet and no bias voltage.
Figure 2 shows the polarization spectrum
for 1) no AHC, 2) 30 minutes AHC, and 3) 60 minutes AHC. All three data
sets are consistent within the statistical errors. No depolarization
was observed after 60 minutes, which is $\sim$100 times longer exposure
time than the optimum time, of AHC.

\begin{figure}[ht]
\centerline{\epsfxsize=2.5in\epsfbox{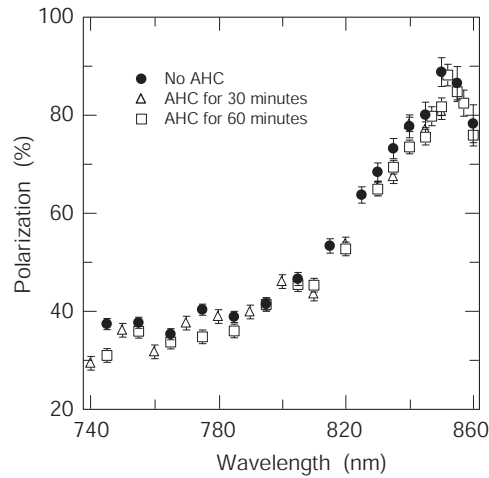}}
\caption{Polarization as a function of wavelength. Solid circles are
measurements without AHC, triangles are measurements after 30 minutes of AHC,
and squares are measurements after 60 minutes of AHC.}
\end{figure}

\section{Conclusions}
Atomic hydrogen cleaning can be used to prepare
high-QE GaAs photocathodes at the lower heat-cleaning temperature of
450$^\circ$C. Photocathode quantum efficiencies as high as 15\% were
obtained when hydrogen ions were eliminated. Extended exposure to atomic
hydrogen was found to have no effect on the electron polarization.

We thank Matt Poelker of Jefferson Lab for providing useful information about
the AHC technique used at his laboratory.

\end{document}